\documentstyle[12pt]{article}
\begin{document}
\title{Output from Bose 
condensates in tunnel arrays: 
the role of mean-field interactions and of transverse confinement}

\author{M.~L. Chiofalo $^*$
and M.~P. Tosi
\\
{\small \it Istituto Nazionale di Fisica
della Materia and Classe di Scienze,}\\
{\small \it  Scuola Normale Superiore, 
I-56126 Pisa, Italy}\\
}

\date{}
\maketitle

\begin{abstract}
We present numerical studies of atomic transport in 3D and 
1D models for a mode-locked, pulsed atom laser as realized 
by Anderson and Kasevich 
[Science 281 (1998) 1686] using an elongated
Bose condensate of ${}^{87}$Rb atoms poured into a vertical optical
lattice. From our 3D results we ascertain in a quantitative manner the
role of mean-field interactions in determining the shape and the size
of the pulses in the case of Gaussian transverse confinement. By
comparison with 1D simulations we single out a best-performing 1D
reduction of the mean-field interactions, which yields quantitatively
useful predictions for all main features of the matter output. 

\end{abstract}

\bigskip
\noindent $^*$ Corresponding author\\
Maria Luisa Chiofalo\\
Scuola Normale Superiore\\
Piazza dei Cavalieri 7, I-56126 Pisa (Italy)\\
Tel.: +39-050-509.058\\
e-mail: marilu@ducky.sns.it

\section{Introduction}

The dynamics of Bose-Einstein condensates of alkali vapours
\cite{cornell1}-\nocite{ketterle0}\cite{hulet}
in very elongated traps 
is a matter of wide experimental and theoretical interest. 
An example is the mode--locked, 
pulsed atom laser which has been realized 
by Anderson and Kasevich \cite{andersonkasevich}
by pouring a condensate of ${}^{87}$Rb atoms
in a vertical optical lattice. Drops of coherent matter leave the
condensate under the effect of gravity. In this and other similar
situations a theorist would like to reduce the problem to
one-dimensional (1D) transport along the axial direction, keeping
account of the interactions through a renormalization of the
scattering length embodying the transverse confinement of the 3D
condensate.

A numerical study of atomic  
transport in a model relevant to the above-mentioned 
system  
has been based on the solution of 
a 1D time-dependent Gross-Pitaevskii equation 
(GPE) \cite{PLA}. This was obtained by freezing out
the transverse motions of the
condensate and by renormalizing the mean-field 
interactions according to a proposal made
by Jackson {\it et
al.} \cite{pethick}. The main result of this study was to show
that, 
independently of the strength of the interactions and in complete
agreement with the experimental data of Anderson and
 Kasevich \cite{andersonkasevich}, the separation between successive
matter drops is determined by  
the period of Bloch oscillations of the condensate in the 1D periodic
potential of the optical lattice. It was further seen from the
numerical results that 
the shape of the drops is closely 
related to that of the parent condensate, thus supporting a mechanism
of coherent emission and suggesting a
practical way to tailor matter-wave laser pulses. Nevertheless, the
question remains of how far a specific 1D schematization may
quantitatively capture other features of the phenomenon.

In this Letter we address this question by 
carrying out a 3D numerical study of gravity-driven transport in a
cylindrically symmetric condensate inside an optical lattice and by
testing against these data the results of 1D reductions of the
problem. In addition to the proposal of 
Jackson {\it et
al.} \cite{pethick}, we test a reduction previously proposed for
condensates in anisotropic harmonic traps
\cite{jamie,schneider}, in which we fix the effective 1D scattering
length by adjusting the chemical potential of the  1D 
model to that of
the 3D condensate. As a further proof that the phenomenon of drop
emission reflects the Bloch oscillations of the condensate, the
spacing between the drops remains the same in all these models. The
focus of our study then is on the shape and size of the matter 
pulses.  

In our numerical simulations we use a fast, explicit time-marching
scheme for the solution of the GPE in cylindrical geometry, which has
been developed by Cerimele {\it et al.} \cite{cps}. This 
method is briefly described in Section 2. Our extensive results of 3D
simulation are presented in Section 3 and summarized in a simple
diagram reporting the fractional number of particles in 
the first drop
as a function of a single, suitably defined combination of system
parameters. The corresponding results from the 1D reductions are
reported in Section 4 and critically compared with those of the 3D
simulation in Section 5. This final section also present our
conclusions.

\section{The numerical method}

The time-dependent GPE for the
condensate wave function $\Psi({\bf r},t)$ is ~\cite{gp,pita}
\begin{equation}
i\hbar{\partial\Psi({\bf r},t)\over\partial t}=
\left(-{\hbar^2\nabla_{\bf r}^2\over 2
M}+U_{ext}({\bf r})+
U_I|\Psi({\bf r},t)|^2\right)\Psi({\bf r},t)\quad ,
\label{gpe}
\end{equation}
where $M$ is the atomic mass, $U_I={4\pi\hbar^2 aN/
M}$ is the interaction strength and
$U_{ext}$ is the external potential,
$a$ being the scattering length and $N$ the number of particles in
the condensate. In what follows $U_{ext}(\mathbf{r})$ is due to 
the optical lattice and to gravity, namely
\begin{equation}
U_{ext}(r,z)=U_l^0[1-\exp(-r^2/r^2_{lb})\cos^2(2\pi z/\lambda)]
-Mgz\; ,
\label{uext}
\end{equation} 
Here, $U_l^0$ is the well depth, 
$r_{lb}$ is the transverse 
size, the wavelength $\lambda$ yields
the lattice period $d=\lambda/2$ and $g$ is the acceleration of
gravity. Finally, 
the normalization condition is $\int
|\Psi({\bf r},t)|^2 d{\bf r} =1$.

We numerically solve Eq. (\ref{gpe}) 
by using an {\it explicit} time marching technique \cite{cps}, 
in contrast with 
alternating-direction-implicit solver
methods previously used in the context of Bose-Einstein
condensation \cite{holland1,prl}. 
The present algorithm 
extends a fast time-staggered
scheme proposed by Visscher \cite{visscher} to solve
the Schr\"odinger equation in an external potential, with the aims of
preserving 
norm conservation in 
the presence of non-linear mean-field 
interactions and of handling cylindrical symmetry \cite{PLA,cps}. 
In brief, in solving Eq. (\ref{gpe}) we 
synchronously advance the real
and imaginary parts of the scaled wave function
in units of two time steps, using their intermediate centred value.
The space derivatives are approximated by using centred
differentiation. 

The following crucial points deserve special comment.
It is proven a priori
\cite{cps} and numerically 
verified that the algorithm preserves the norm 
of the wave function at each time step, provided that the boundary 
conditions are such as to annihilate surface terms.
It can also be shown \cite{cps} that the numerical stability  
of the algorithm is preserved as long as 
the simulation time-step does not exceed a critical value 
$\Delta\tau_c$, which 
is limited either by the grid spacing or by the magnitude 
of the product $aN$ entering $U_I$. In our simulations the actual
time-step is consistently kept well below the marginal stability
threshold. Finally,
the cylindrical symmetry is handled by an accurate
treatment of the space derivatives near the symmetry axis 
\cite{holland1}.

\section{Results for a 3D condensate in cylindrical geometry}
Equation (\ref{gpe}) is made dimensionless 
by adopting the scale units
$S_l=\sqrt{\hbar/2M \omega}$, $S_t=1/\omega$ and 
$S_E=\hbar \omega$
for length, time and energy,
$\omega$ being the radial frequency of the original 
magnetic trap. We also rescale the wave function
by the dimensionless radial
coordinate $\rho$. 
We use a grid resolution of $21\times 16$ on each single well, 
the time step being $\Delta\tau=2\cdot 10^{-6}$. 

The initial value $\Psi(r,z;t=0)$ of the wave function
is chosen so as to approximate the condensate realized in 
\cite{andersonkasevich}. Namely,
\begin{equation}
\label{initcond}
\Psi(r,z;t=0)=A\exp [-M(\omega_{r}
r^2+\tilde{\omega} z^2)/2\hbar]\sum_l
\exp [-M\omega_{z} (z-ld)^2/2\hbar]\; .
\end{equation}
Here, $A$ is a normalization factor and 
$l$ labels the occupied sites, their total number being $n_w$. 
In constructing Eq. (\ref{initcond}) 
we have assumed (i) constant phase of the condensate in space
\cite{band}; (ii) a Gaussian transverse profile with a frequency 
$\omega_{r}=\sqrt{2 U_l^0/(M r^2_{lb})}$ given by 
the harmonic approximation to the transverse shape of the
optical potential; and (iii) an overall axial profile reflecting 
that 
of the condensate inside the 
magnetic trap before loading the optical lattice and 
taken as
a Gaussian having frequency 
$\tilde{\omega}=4\pi^{3/5}/\zeta^2\omega $, 
with $\zeta=(32\pi Na/S_l)^{1/5}$  
\cite{baym}. Finally, the lowest state at each lattice site is
occupied by a portion of condensate, having Gaussian shape with
frequency $\omega_z=2\sqrt{U_l^0 E_R}/\hbar$, where
$E_R=h^2/2M\lambda^2$ is the recoil energy.

Before entering the presentation of the simulation results, we list
below 
the system parameters relevant to the experiment on ${}^{87}Rb$
\cite{andersonkasevich}. These are
$a=110 a_0$ with $a_0$ the Bohr
radius, $N=10^4$, $\lambda=850\; nm$, $r_{lb}=80\; \mu m$,
$U_l^0=1.4\;\; E_R$ and $n_w=31$.
The parameters in this reference list 
provide our reference run. 

Fig. 1 shows four
pictures of drop emission for different values of the
coupling strength, 
by plotting 
the contour density profiles, 
taken with $U_l^0=1.4\; E_R$
after 5.3 $ms$. The first panel displays
the behaviour of the non-interacting 
gas, namely the case $a=0$ and  $n_w=31$. The other panels 
report the behaviour of the interacting gas with $a=110\;
a_0$; from left to right, 
the cases  $N=10^4$ with $n_w=31$, 
$N=10^5$ with $n_w=49$ and $N=2\cdot 10^5$ with $n_w=57$. 
All the other parameters are 
as listed above. 

A number of the results obtained in the earlier 1D simulation 
\cite{PLA} are recovered in the present 3D runs. Each drop in 
figure 1 
extends over a number of
wells equal to that occupied by the parent
condensate. In all cases the drops are equally 
spaced by seventy wells from centre to centre. This spacing 
corresponds to 1.1 $ms$ of simulation time,
in agreement with experiment \cite{andersonkasevich} and
with the value of 1.09 $ms$ for the period 
$T_B=2h/Mg\lambda$ of Bloch oscillations of the condensate in the
periodic optical potential. The time lag between successive drops is
independent of the amplitude of the periodic potential, 
of the strength of the interactions and of the dimensionality of the
simulation sample, as expected if Bloch oscillations provide the
correct interpretation of the observations \cite{band}.
Finally, both the axial width and the fine structure of each drop 
reproduce those of the parent condensate, as expected for coherent
emission from all lattice wells.

Again in analogy with the case of 1D simulation \cite{PLA}, the
transport behaviour of the 3D system as a function of its governing
parameters can be summarized in a single diagram. We introduce a
scaling parameter $g_s$, having the dimensions of the acceleration of
gravity, by setting
\begin{equation}
Mg_s d=U_l^0-U_i
\label{gs}
\end{equation}
where $U_i\equiv 4\pi\hbar^2 a N/M R^3$ with 
$R=\zeta S_l$ is a measure of the mean-field 
interaction strength. In figure 2 we plot
the fractional number 
$N_{drop}/N$ of atoms in the first drop
as a function of the ratio $g/g_s$, each symbol representing a 
run with different input parameters.
Symbols
of different shape represent runs at different values of 
$U_l^0/E_R$, as
shown in the legend. For each symbol we report the
values of $N_{drop}/N$ 
corresponding to increasing coupling strength $U_i$, 
starting from the non-interacting 
gas and then increasing $N$ 
in the sequence $N=10^4$, $10^5$ and 
$2\cdot 10^5$ (from left to right).

It is seen from figure 2 that there is a
critical value for the onset of drop formation, which is 
$g/g_s\simeq 0.14$ (the onset is marked by an arrow).
To all effects there is no emission of drops for
subcritical values of $g/g_s$, since the 
condition of resonance between the bound state in the well
and the continuum is not satisfied. 
Well defined drops are instead emitted at
supercritical values of $g/g_s$. 
In this regime $N_{drop}/N$ 
increases rather regularly with
$g/g_s$ and shows little sensitivity to the strength of the 
mean-field interactions at fixed $U_l^0$. Ultimately, 
with decreasing $U_l^0$ the potential wells become too 
shallow and regular drop emission turns into a discharge 
of the whole condensate.

On the other hand, 
an increase in the mean-field interactions affects 
the axial width of the drops. In particular the
width of the first drop, as measured by its second moment, 
is close
to $n_w/2$, with an appreciable scatter from case to case, while
the centre-to-centre distance of neighbouring drops is 
about 70 wells. As a result, overlap 
between drops starts at
$N \simeq 3\cdot 10^5$.

In summary, our 3D simulations confirm that the 
transport behaviour of
a Bose condensate in a vertical optical lattice is described by a
diagram as shown in figure 2. This diagram could yield useful
predictions even for atomic species different from 
$^{87}$Rb.

\section{One-dimensional modelling }
The time-dependent GPE for a 1D reduction of the present transport
problem is \cite{PLA}
\begin{equation}
i\hbar{\partial\psi({z},t)\over\partial t}=
\left(-{\hbar^2\nabla_{z}^2\over 2
M}+u_{ext}({z})+
u_{I}|\psi({z},t)|^2\right)\psi({z},t)\quad ,
\label{gpez}
\end{equation}
where $u_{ext}(z)=U_l^0\sin^2(2\pi z/\lambda)-Mgz$ and 
$u_{I}=4\pi\hbar^2\tilde{a}N/M$, 
$\tilde{a}$ being a renormalized
coupling parameter with the dimensions of an inverse 
length.

In the renormalization proposed 
by Jackson {\it et al.} \cite{pethick} (hereafter referred to as
model I) one assumes that the coherence length of the condensate in
the axial direction is much larger than its transverse radius. The 3D
wave function is factorized as $\Psi({\mathbf r},t)=
g(r,\sigma(z(t)))$
where $\sigma(z)$ is the axial density.
Using a harmonic
approximation for the radial part of the 
optical potential, one obtains in the
specific problem an effective scattering length
$\tilde{a}^{I}\equiv\gamma^I a$ with 
$\gamma^I= \sqrt{U_l^0/E_R}/(r_{lb}\lambda)$. 
A similar renormalization
of $U_l^0$ is negligible,
since $r_{lb}\gg\lambda$. 

As an alternative (hereafter referred to as model II)
we impose that the value of the 
chemical potential of the 3D
system be preserved upon reduction to 1D \cite{jamie,schneider}.
As is shown in a
detailed calculation given in Appendix A
within the Thomas-Fermi approximation, we find
$\tilde{a}^{II}=\gamma^{II}a$ with $\gamma^{II}$ given (for
$U_l^0>\mu$) by 
\begin{equation}
\gamma^{II}={1\over \pi r_{lb}^2}+
{U_l^0-\mu\over 4 E_R}
{1\over \lambda aN}I(\mu)\; ,
\label{gamma}
\end{equation}
an explicit expression for the positive quantity 
$I(\mu)$ being given in the Appendix. We remark for later discussion
that $\gamma^{II}$ in Eq. (\ref{gamma}) depends implicitly on the
product $aN$ which determines 
the interaction strength.

Table 1 reports the values of $\gamma^I S_l^2$ and 
$\gamma^{II} S_l^2$ for a number of values of 
$U_l^0/E_R$ and of $N$, all other
parameters being as in the reference list. The 
values of $\mu/E_R$ are also shown. 
The important point of Table 1 is that while $\gamma^{I}$ is
independent of  $N$, $\gamma^{II}$ 
decreases with increasing $N$. This means that in 
model I an increase 
in the number of particles leads to a much more rapid increase 
of the mean-field interaction parameter.
For instance, in the case $U_l^0/E_R=1.4$
an increase of $N$ by a factor $30$ implies that 
$\gamma^{II}aN$ increases by only 
a factor $6.8$.     

In the next Section we discuss the consequences 
of these different behaviours of the two 1D models. We
compare the results of the 3D simulation
reported in Section 3 with those 
obtained by solving the 1D
GPE with $\tilde{a}^I$ or 
$\tilde{a}^{II}$ inserted in turn into the mean-field term $u_I$. 

\section{Discussion and concluding remarks}

Table 2 collects the results for the fractional number 
$N_{drop}/N$ of atoms in the first drop
for different values of $U_l^0/E_R$ and $N$ in the interacting 
gas, as calculated from 
(i) the 3D simulation in cylindrical geometry 
(third column), (ii) the 1D simulation 
in model I (fourth column), and (iii) the 1D simulation in model II
(last column).

It is immediately seen that model II works better than model I
and quantitatively reproduces the data of the 3D simulation. 
The interactions lift the bound state 
towards the
continuum by an amount which may be measured by the
mean interaction energy $E_{I}$ per particle . This is 
proportional to the product of the effective scattering 
length times the particle density. In 
the 1D simulation according to model I we have 
$E_I\propto \tilde{a}/\lambda\propto a/\lambda^2 
r_{lb}$. 
In the 3D simulation we have instead $E_I\propto a/(\lambda
r^2_{lb})$, which is significantly smaller since
$r_{lb}\gg \lambda$.

Thus a picture emerges in which axial transport is accompanied by a
transverse breathing of the condensate, due to the vanishing 
of radial
confinement at $z=(2n+1)\lambda/4$ (see Eq. (\ref{uext})). 
Model I cannot account for this behaviour,
since it has been derived by freezing the transverse 
motion.  
It also
follows that 
an increase in coupling strength is partially accomodated
by a transverse spreading of the condensate, so that  
the value of $N_{drop}/N$ is rather insensitive to 
the mean-field 
interactions in the present range of system parameters.

In summary, a Bose condensate in an optical lattice, subject to a
constant driving field in the linear transport regime, behaves as a
coherent blob of matter which executes Bloch oscillations through 
band
states and can undergo Zener tunnel at the Brillouin zone edge. 
The simple diagram shown in Figure 2 describes the drops of coherent
matter which are emitted {\it via} tunnel into the continuum, in
dependence of the governing parameters of the system. The effect of
the mean-field interactions is very small within the range of system
parameters in which regular pulses are emitted. A quantitative
reduction of this behaviour to a 1D model can be achieved by imposing
a simple condition of constancy of the chemical potential. We expect
that this method of dimensionality reduction will be useful in 
other similar problems and applications.

\bigskip

\noindent{\bf Acknowledgements}

\noindent We gratefully thank Dr. M. M. Cerimele, Dr. F. Pistella and
Professor S. Succi of the Istituto Applicazioni Calcolo ``M. Picone''
of the Italian National Research Council for making the simulation
programs available to us and for useful discussions. Special thanks
are due to Dr. F. Pistella for her help with the graphics. This work
has been sponsored by the Istituto Nazionale di Fisica della Materia
under the Advanced Research Project on BEC.

\appendix
\catcode `\@=11
\@addtoreset{equation}{section}
\def\theequation{A.\arabic{equation}}
\catcode `\@=12
\section{Calculation of the renormalization factor $\gamma^{II}$}
We set $\Psi({\mathbf r},t)=\exp (i\mu t/\hbar)\Psi({\mathbf r})$ 
in Eq. (\ref{gpe}) and 
$\psi(z,t)=\exp (i\mu t/\hbar)\psi(z)$ in Eq. (\ref{gpez}). 
We then have to solve the stationary GPE's
\begin{equation}
\mu\Psi ({\bf r})=
\left(-{\hbar^2\nabla_{\bf r}^2\over 2
M}+U_{ext}({\bf r})+
U_I|\Psi({\bf r})|^2\right)\Psi({\bf r})
\label{gpes}
\end{equation}
for $\Psi(\mathbf{r})$ in the 3D and
\begin{equation}
\mu\psi({z})=
\left(-{\hbar^2\nabla_{z}^2\over 2
M}+u_{ext}({z})+
u_{I}|\psi({z})|^2\right)\psi({z})\quad .
\label{gpezs}
\end{equation}
for $\psi(z)$ in 1D. 
To this end we use the Thomas-Fermi approximation,
which amounts to neglecting the kinetic energy terms.
We then 
impose the normalization condition on 
$\Psi({\mathbf r})$ and on $\psi(z)$, thereby obtaining the
expressions for 
the chemical
potential $\mu$.

After introducing dimensionless quantities as indicated in Section 3
and assuming confinement ($U_l^0>\mu$), this procedure yields the
relations 
\begin{equation}
{32\pi^2 aN \gamma^{II}S_l^2\over U_l^0\lambda}=
\left(1-2\beta\right)\arccos\left(2\beta-1\right)+
\sqrt{1-\left(2\beta-1\right)^2}
\label{mucond1}
\end{equation}
in the 1D case and
\begin{equation}
{32\pi^2 aN S_l^2\over U_l^0\lambda}=
{32\pi^2 aN \gamma^{II} S_l^2\over U_l^0\lambda}
-2\beta\int^{\arccos(2\beta-1)}_0 w\tan(w/2) dw
\label{mucond3}
\end{equation}
in the 3D case, where $\beta\equiv(U_l^0-\mu)/U_l^0$. 
These equations yield 
Eq. (\ref{gamma}) in the main text, where 
\begin{equation}
I(\mu)=\int^{f(\mu)}_0 
w\tan(w/2) dw>0
\end{equation}
with
\begin{equation}
f(\mu)=\arccos(2(U_l^0-\mu)/U_l^0-1)\; .
\end{equation}
In our numerical calculations we first determine $\mu$ from
Eq. (\ref{gpes}) 
and then $\gamma^{II}$ from Eq. (\ref{gamma}).

\newpage

\noindent{\bf Figure captions}

\bigskip\noindent {\bf Fig. 1}: Contour plots
of the condensate density after 5.3 $ms$ for $U_l^0=1.4\; E_R$, as
functions of the axial coordinate $z/d$ and of the transverse 
distance in $\mu m$. 
The first panel on the left refers to the
non-interacting gas with a number of particles $N=10^4$. 
The other panels refer to the interacting
gas, for $a=110\; a_0$ and various values of $N$ 
(from left to right,
$N=10^4$, $10^5$ and $2\cdot 10^5$).

\bigskip\noindent {\bf Fig. 2}: 
Diagram for drop formation from 3D simulation in
cylindrical geometry. We plot the fractional number $N_{drop}/N$ 
of particles in the first drop against $g/g_s$
for $g=981$ cm$/$s${}^2$.
The meaning of the symbols is explained in
the legend and in the text.

\newpage
\begin{table}
\centerline{
\begin{tabular}{|c|c||c|c|c|}
\hline
$U_0/E_{R}$&$N$&$\gamma^I S_l^2$&$\gamma^{II} S_l^2$&$\mu/E_R$\\
\hline\hline
0.4
&$10^4$
&$7.9\cdot 10^{-2}$
&$3.4\cdot 10^{-3}$
&0.11
\\
&$10^5$
&
&$1.2\cdot 10^{-3}$
&0.25
\\
&$2\cdot 10^5$
&
&$8.5\cdot 10^{-4}$
&0.31
\\
&$3\cdot 10^5$
&
&$6.8\cdot 10^{-4}$
&0.34
\\
\hline
0.7
&$10^4$
&0.10
&$4.3\cdot 10^{-3}$
&0.16
\\
&$10^5$
&
&$1.6\cdot 10^{-3}$
&0.36
\\
&$2\cdot 10^5$
&
&$1.1\cdot 10^{-3}$
&0.45
\\
&$3\cdot 10^5$
&
&$9.2\cdot 10^{-4}$
&0.51
\\
\hline
1.4
&$10^4$
&0.15
&$5.7\cdot 10^{-3}$
&0.24
\\
&$10^5$
&
&$2.1\cdot 10^{-3}$
&0.57
\\
&$2\cdot 10^5$
&
&$1.6\cdot 10^{-3}$
&0.72
\\
&$3\cdot 10^5$
&
&$1.3\cdot 10^{-3}$
&0.83\\
\hline
2.1
&$10^4$
&0.18
&$6.8\cdot 10^{-3}$
&0.31
\\
&$10^5$
&
&$2.6\cdot 10^{-3}$
&0.73
\\
&$2\cdot 10^5$
&
&$1.9\cdot 10^{-3}$
&0.94
\\
&$3\cdot 10^5$
&
&$1.6\cdot 10^{-3}$
&1.08
\\
\hline
\end{tabular}
}
\caption{
Calculated values of $\gamma^I S_l^2$, 
$\gamma^{II} S_l^2$ and $\mu/E_R$ for various values of 
$U_l^0/E_R$ and $N$.}   
\end{table}  

\begin{table}
\centerline{
\begin{tabular}{|c|c||c|c|c|}
\hline
$U_0/E_{R}$ & $N$ &\multicolumn{3}{c|}
{$N_{drop}/N$}\\
\hline
& & 3D &1D Model I& 1D Model II\\
\hline\hline
0.7
&$10^4$
&0.61
&0.60
&0.60
\\
&$10^5$
&0.61
&0.60
&0.61
\\
&$2\cdot 10^5$
&0.61
&0.66
&0.62
\\
\hline
1.4
&$10^4$
&0.12
&0.12
&0.12
\\
&$10^5$
&0.11
&0.26
&0.11
\\
&$2\cdot 10^5$
&0.11
&0.32
&0.10
\\
\hline
2.1
&$10^4$
&0.01
&0.01
&0.01
\\
&$10^5$
&0.01
&0.06
&0.01
\\
&$2\cdot 10^5$
&0.007
&0.02
&0.007
\\
\hline
\end{tabular}
}\caption{
Fractional number 
$N_{drop}/N$ of atoms in the first drop
for various values of $U_l^0/E_R$ and $N$.}   
\end{table}  
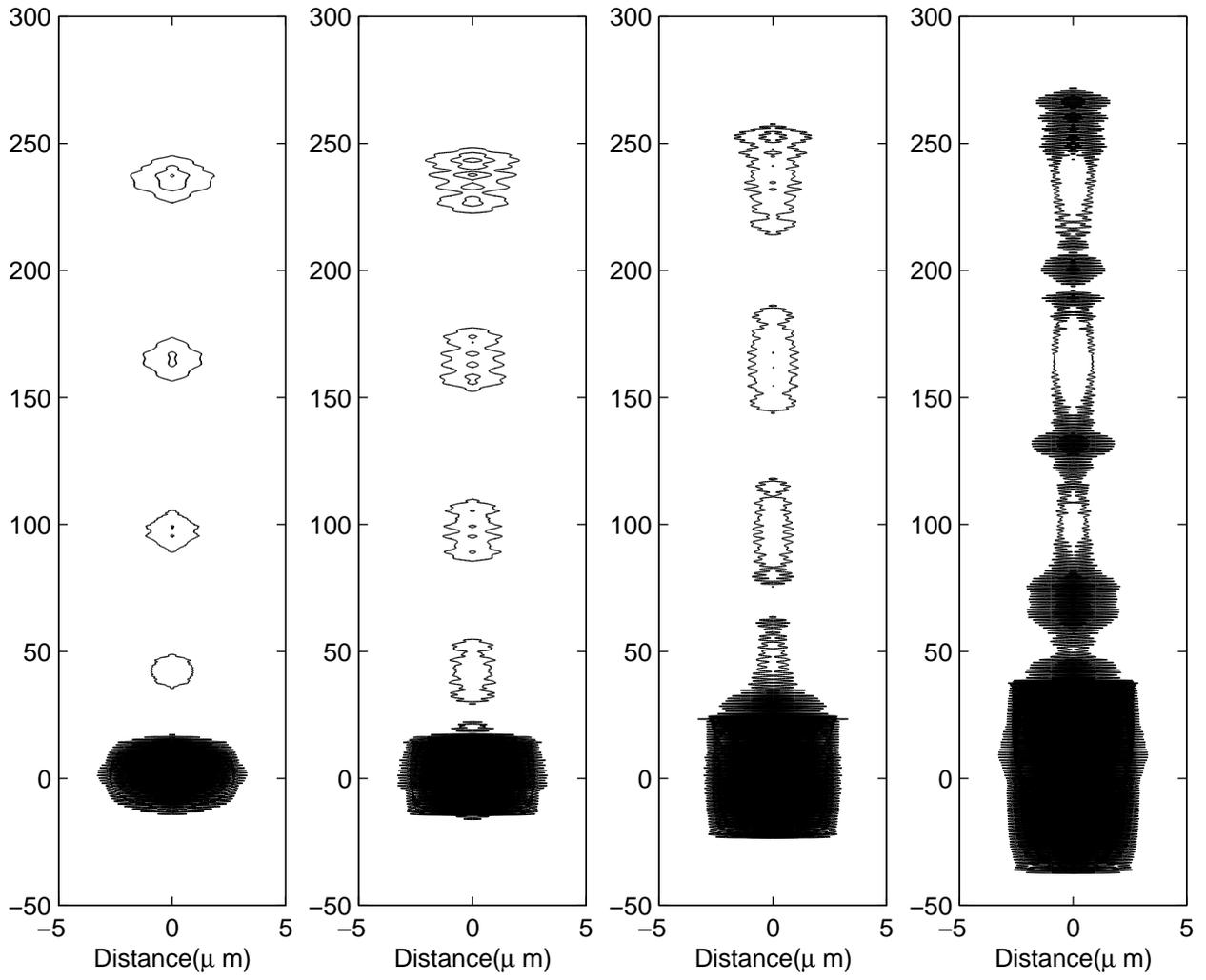
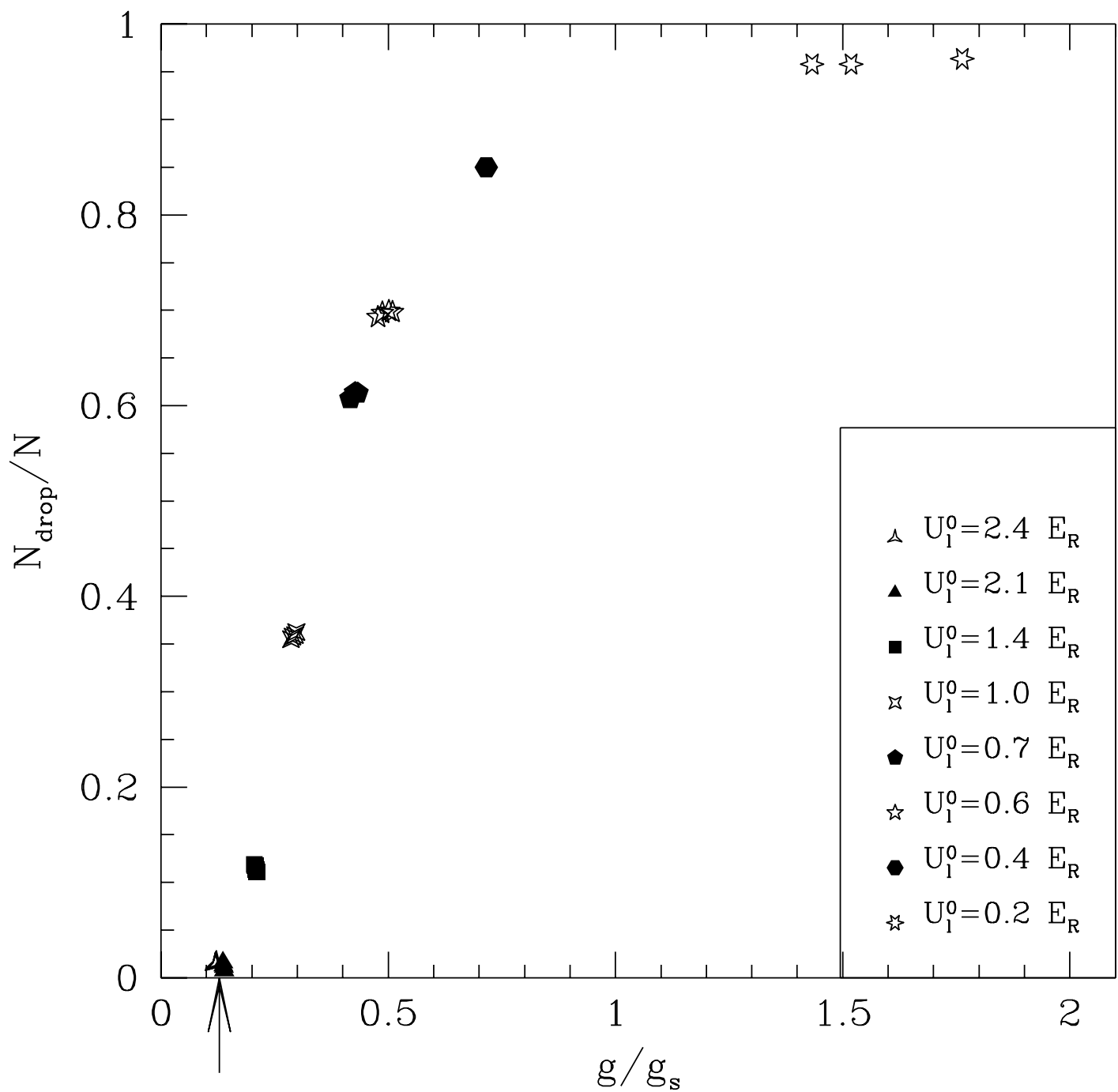
\end{document}